\documentclass{article}
\usepackage{spconf,amsmath,graphicx,hyperref}
\usepackage{amssymb}
\title{Beyond Answer Changes: Matched Same-Class Controls\\
for Local Audio Evidence}
\name{Yunuo Wang$^{1,\dagger}$ \qquad Maoyuan Chen$^{2,\dagger}$ \qquad Yuanjin Huang$^{3}$ \qquad Xing Liu$^{4}$ \thanks{$^{\dagger}$These authors contributed equally to this work.}}
\address{\parbox[t]{\dimexpr\textwidth-2\tabcolsep\relax}{\centering
$^{1}$The Hong Kong University of Science and Technology (Guangzhou), Guangzhou, China\\
$^{2}$School of Computer Science, Wuhan University, Wuhan, China\\
$^{3}$National University of Singapore, Singapore\\
$^{4}$Northeastern University, Toronto, Canada}}
\begin{document}
\maketitle
\begin{abstract}
For audio-language models, an answer change after a local sound-event edit can reflect a response to the event's class, sensitivity to waveform rewriting, or both. We compare matched same-class and different-class replacements of a queried event. Both use the same interval, level estimate, donor fitting, and crossfade, while the same-class replacement preserves the correct answer. Across 600 synthetic soundscapes and three models from two architecture families, different-class replacements produce 49.7 to 60.9 percentage points more answer changes than same-class replacements. Same-class accuracy remains within 5.2 percentage points of original accuracy. Among trials with correct original and same-class answers, soundscape-weighted selective accuracy for the inserted class is 73.7 to 81.9\%, averaging eligible trials within each soundscape and then weighting eligible soundscapes equally. This protocol tests how model answers respond to local changes in event class.
\end{abstract}

\begin{figure*}[t]
\centering
\includegraphics[width=\textwidth]{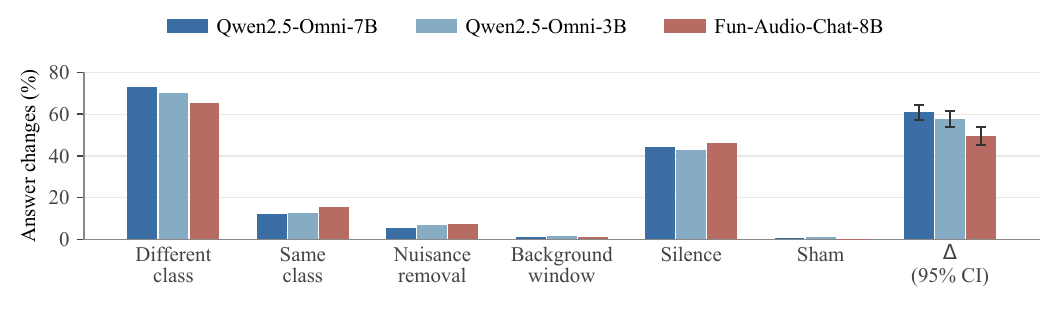}
\caption{Changing the queried event's class elicits more answer change than
rewriting the same interval with another recording of that class. Nuisance and
sham edits produce smaller responses. The rightmost group shows the paired
gap $\Delta$ and its 95\% interval. The bootstrap resamples
soundscapes with both option orders and all conditions kept together.}
\label{fig:conditions}
\end{figure*}

\section{Introduction}
\label{sec:intro}

Audio-language models answer natural-language questions about soundscapes by
combining sound recognition with language-based reasoning
\cite{qwen25omni2025,funaudiochat2026}. Benchmarks such as the Massive Multi-Task
Audio Understanding and Reasoning benchmark span speech, environmental sounds,
and music \cite{mmau2024}. A central question is whether an answer is genuinely
supported by the acoustic evidence. Correct responses can often be recovered
from text alone \cite{authorship2026impact}, and improvements in audio question
answering can partly reflect better text-based reasoning \cite{omnir1}. For
environmental sounds this dependence is also local, since a question about a
brief bark concerns one event within a longer recording. Evaluating the answer
therefore requires linking it to the class of that event.

Intervention methods examine evidence use by perturbing the audio and observing
how the answer moves. Noise and region-guided masking probe audio faithfulness
\cite{lalmfaith2026}. Attention-guided masking with a duration-matched control
connects localized perturbations to prediction changes \cite{mapo2026}.
Counterfactual voice grounding tests speaker-attributed reasoning
\cite{hear2026whosaidwhat}, and paralinguistic audits vary acoustic cues while
holding transcripts fixed \cite{paraling2026}. These approaches provide useful
tests of acoustic grounding. For a local environmental-sound question, an answer
change depends on what the intervention changes at the queried location. A
replacement can change the event's class or rewrite the same interval with
another recording of that class. The matched same-class replacement measures
responses to local rewriting while preserving the queried class. Comparing it
with a different-class replacement asks how much additional answer movement
comes from changing that class under the same editing procedure. Removal and
masking tests instead probe what happens when local evidence is deleted or
degraded. The same
concern arises in text-domain counterfactual prompting, where a targeted edit
is a compound treatment that benefits from a meaning-preserving baseline
\cite{yang2026compared}.

This paper addresses how to attribute the movement of an answer after a local
waveform intervention. \emph{When an audio-language model changes its answer
after a local waveform intervention, how can we tell whether it responded to
the event's class or to the fact that the waveform was edited?} We introduce a
matched local waveform intervention protocol on synthetic soundscapes with
annotated event intervals, and make three contributions.
(i)~\textbf{A matched replacement protocol}. Class-changing and class-preserving
replacements share the target span, donor-fitting procedure, level estimate,
and crossfade. The class-preserving arm measures response to local waveform
rewriting. Nuisance, background-window, silence, and sham controls complete the
design.
(ii)~\textbf{A measure of answer changes and their direction}. We report
whether the answer changes and whether it identifies the replacement class.
An exact decomposition relates the observed gap to an accuracy-based
independence reference and covariance between condition outcomes.
(iii)~\textbf{A cross-architecture behavioral result}. Across Qwen2.5-Omni-7B
and Qwen2.5-Omni-3B \cite{qwen25omni2025} and Fun-Audio-Chat-8B
\cite{funaudiochat2026}, class-changing replacements produce 49.7 to 60.9
percentage points more answer changes than matched class-preserving rewrites.
Sham answer changes stay below 1.0\%, and soundscape-weighted selective
accuracy reaches 73.7 to 81.9\%.

\section{Soundscape intervention protocol}
\label{sec:protocol}

Figure~\ref{fig:conditions} previews the response profile of the interventions.
Each soundscape contains a localized query event, a temporally separate
nuisance event, and continuous ambience. One of the two answer options names
the query class. The other names a class absent from the placed foregrounds,
and neither option names the nuisance event.

The different-class variant replaces the query event with a recording from the
alternative class. Its matched same-class control inserts a different recording
of the query class into exactly the same interval. Both variants use the same
donor-fitting, level-estimation, background-copying, and crossfade procedures.
Their comparison measures the response to changing the queried class relative
to ordinary rewriting of that interval.

Donor segments remain within their annotated event intervals, and separate
foreground events are at least 0.3\,s apart. The same-class and different-class
replacements use the same target interval and matched edit durations. Post-render
checks verify a median background-noise-floor difference below 1\,dB between
replacement and original spans, fewer than 5\% of soundscapes above a 10\,dB
difference, and boundary first differences within three times those of the
original.

The original audio anchors the six interventions. The sham self-replacement
follows the same editing path, with floating-point rounding changing at most
1 LSB in about 6 of 160,000 samples. The same-class rewrite preserves the
queried class, and whole-clip silence removes the soundscape. The different-class
rewrite substitutes an event from the alternative class. Nuisance removal and
an event-free background-window rewrite test edits away from the query.
Each waveform is queried in both option orders with
``Which of these two sounds is present in it?'', requesting
\texttt{Answer: A} or \texttt{Answer: B}.

\begin{table*}[t]
\centering
\caption{Answer changes relative to the original answer (\%) across 1,200
trials per model. Each of 600 soundscapes contributes two option orders and six
interventions. $\Delta$ is the different-class minus same-class answer-change
gap. The 95\% bootstrap intervals resample paired soundscapes, keeping both
option orders and intervention conditions together.}
\label{tab:conditions}
\begin{tabular}{lrrrrrrr}
\hline
Model & Different & Same & Nuisance & Background- & Silence & Sham &
$\Delta$ (95\% CI) \\
 & class & class & removal & window & & & \\
\hline
Qwen2.5-Omni-7B & 73.17 & 12.25 & 5.50 & 1.33 & 44.42 & 0.58 &
\textbf{60.92} [57.25, 64.58] \\
Qwen2.5-Omni-3B & 70.25 & 12.50 & 7.08 & 1.50 & 42.75 & 0.92 &
\textbf{57.75} [54.00, 61.67] \\
Fun-Audio-Chat-8B & 65.33 & 15.67 & 7.50 & 1.08 & 46.00 & 0.17 &
\textbf{49.67} [45.33, 53.92] \\
\hline
\end{tabular}
\end{table*}

\section{Answer changes and selective accuracy}
\label{sec:metrics}

\begin{figure}[t]
\centering
\includegraphics[width=\columnwidth]{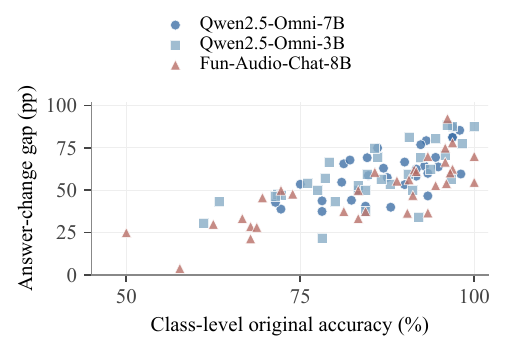}
\caption{Class-level original accuracy versus the different-class minus
same-class answer-change gap. Each point represents one of 33 query classes
for one model. Both quantities use the same evaluation trials.}
\label{fig:class}
\end{figure}

The matched same-span control gives the response to rewriting an event while
preserving its class. To interpret the contrast with a different-class edit,
let $C_o$, $C_d$, and $C_s$ denote binary correctness on the original,
different-class, and same-class inputs, with $a_j=\mathbb{E}[C_j]$. For valid
binary answers with fixed option-to-class mappings, the original and
same-class inputs share a truth label. The different-class input has the other
label. Their answer-change probabilities satisfy
\begin{equation}
\begin{aligned}
A_s &= \mathbb{E}[C_o+C_s-2C_oC_s],\\
A_d &= \mathbb{E}[1-C_o-C_d+2C_oC_d].
\end{aligned}
\label{eq:binary-change}
\end{equation}
The answer-change gap $\Delta=A_d-A_s$ combines marginal correctness and
dependence between condition outcomes. Factoring the correctness products
gives the reference value under independence of original correctness from each
edited outcome,
\begin{equation}
G_{\mathrm{ind}}=(2a_o-1)(a_d+a_s-1).
\label{eq:independent-reference}
\end{equation}
The exact relationship is
\begin{equation}
\Delta=G_{\mathrm{ind}}+2\operatorname{Cov}(C_o,C_d)
  +2\operatorname{Cov}(C_o,C_s).
\label{eq:dependence-correction}
\end{equation}
The reference assumes independence of original correctness from each edited
outcome, without requiring independence between the edited outcomes. It is an
algebraic benchmark based on marginal accuracies. Equation~\ref{eq:dependence-correction}
shows that the difference between the observed gap and this reference equals
the covariance correction. Figure~\ref{fig:class} relates original accuracy
to the gap for each evaluated query class and model.

Selective accuracy measures whether a revised answer identifies the inserted
class on eligible trials. An eligible trial has correct answers on both the
original and same-class inputs. Each soundscape $i$ contributes two option-order
trials $\mathcal T_i$. Define the eligible trials $\mathcal C$ and soundscapes
$\mathcal I_{\mathcal C}$ containing at least one eligible trial. The reported
selective accuracy is
\begin{equation}
\begin{aligned}
\mathcal C &= \{t\mid C_o(t)=C_s(t)=1\},\\
\mathcal I_{\mathcal C} &= \{i\mid \mathcal T_i\cap\mathcal C\ne\varnothing\},\\
F_{\mathcal C} &= \frac{1}{|\mathcal I_{\mathcal C}|}
\sum_{i\in\mathcal I_{\mathcal C}}
\frac{\sum_{t\in\mathcal T_i\cap\mathcal C}C_d(t)}
{|\mathcal T_i\cap\mathcal C|}.
\end{aligned}
\label{eq:selective-accuracy}
\end{equation}
On an eligible trial, identifying the different-class donor also means
changing the original answer in the correct direction. Eligible trials are
averaged within each soundscape before eligible soundscapes receive equal
weight. Eligibility depends on each model's responses, so a soundscape may
contribute one option order. Table~\ref{tab:follow} reports eligible-trial and
eligible-soundscape counts as coverage.

\begin{table*}[t]
\centering
\caption{Correct identification before and after replacement. Selective
accuracy averages eligible trials within each soundscape and then weights
eligible soundscapes equally. Eligibility requires correct original and
same-class answers. The first three accuracy columns use all 1,200 trials.
Eligible trial and soundscape counts summarize coverage.}
\label{tab:follow}
\setlength{\tabcolsep}{6pt}
\begin{tabular}{lcccccc}
\hline
Model & Original & Same-class & Different-class & Selective & Eligible &
Eligible \\
 & accuracy & accuracy & accuracy & accuracy & trials & soundscapes \\
\hline
Qwen2.5-Omni-7B~\cite{qwen25omni2025} & .882 & .859 & .845 & .819 & 971 & 542 \\
Qwen2.5-Omni-3B~\cite{qwen25omni2025} & .853 & .841 & .838 & .801 & 941 & 544 \\
Fun-Audio-Chat-8B~\cite{funaudiochat2026} & .847 & .795 & .800 & .737 & 891 & 481 \\
\hline
\end{tabular}
\end{table*}

\section{Experiments}
\label{sec:results}

\subsection{Experimental setup}
We evaluate Qwen2.5-Omni-7B and Qwen2.5-Omni-3B
\cite{qwen25omni2025}, alongside Fun-Audio-Chat-8B
\cite{funaudiochat2026}, on 600 shared synthetic soundscapes. Scaper
\cite{salamon2017scaper} combines single-label foreground recordings from
the FSD-MIX-SED source release \cite{wang2021shots}, derived from FSD50K
\cite{fonseca2021fsd50k}, with backgrounds from TUT Urban Acoustic Scenes
2018 \cite{heittola2018tut}. Each 10\,s recording sampled at 16\,kHz holds a
query event, a separate nuisance event, and continuous ambience at a nominal
$+10$\,dB event-to-background ratio. The evaluation covers 33 query classes
with event instances held out from the pilot. Two option orders yield 1,200
trials per condition and model. All three models exceed the protocol's 0.75
original-accuracy floor, and silence changes more than 20\% of answers.
Confidence intervals resample paired soundscapes, keeping both option orders
and all intervention conditions together.

\subsection{Class changes versus matched rewrites}
Different-class replacement changes the intended answer, while same-class
replacement rewrites the queried interval and preserves that answer.
Figure~\ref{fig:conditions} shows the resulting separation. The paired
answer-change gaps are 0.609, 0.578, and 0.497 for Qwen2.5-Omni-7B,
Qwen2.5-Omni-3B, and Fun-Audio-Chat-8B, respectively. Their 95\% intervals
in Table~\ref{tab:conditions} exclude zero. The independence reference in
Eq.~\ref{eq:independent-reference} is 0.538, 0.479, and 0.413. The observed
gaps exceed these values by 0.072, 0.099, and 0.084, which equal the
covariance corrections in Eq.~\ref{eq:dependence-correction}. The direction
holds across both model families and all 33 query classes. The smallest
class-level gaps are 3.8 percentage points for Fun-Audio-Chat-8B and 21.9
points for Qwen2.5-Omni-3B. The same-class response provides a reference for
answer changes caused by local waveform editing.

\begin{figure}[t]
\centering
\includegraphics[width=\columnwidth]{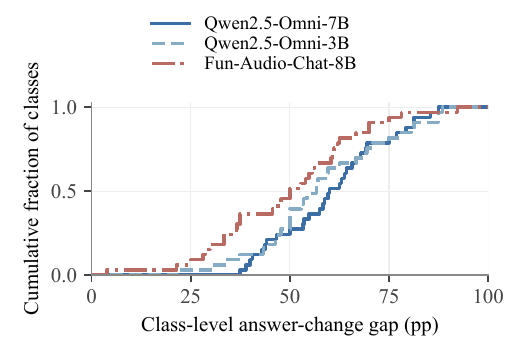}
\caption{Empirical cumulative distributions of class-level answer-change gaps.
Each curve weights the 33 query classes equally and shows the fraction with a
gap at or below each threshold. Gaps are measured in percentage points.}
\label{fig:distribution}
\end{figure}

Figures~\ref{fig:class} and~\ref{fig:distribution} summarize the class-level
results. The scatter relates original recognition accuracy to the gap. The
cumulative curves display its distribution over all 33 query classes with
equal class weights. The positive class-level gaps show that the contrast
extends across all evaluated categories.

\subsection{Locality and editing controls}
Table~\ref{tab:conditions} reports the intervention profile. Nuisance removal
changes answers at rates of 0.055, 0.071, and 0.075 for the three models,
while background-window rates stay below 0.016 and sham rates below 0.010.
The much larger target response thus concentrates sensitivity on the queried
event, beyond the response to off-target edits or to the editing path alone.

\emph{Acoustic distance.} Among the 236 soundscapes, or 472 option-order
trials, with different-class and same-class log-mel edit distances matched
within 15\%, the trial-level gap is 0.617 for Qwen2.5-Omni-7B with a 95\%
interval of [0.555, 0.678]. For Qwen2.5-Omni-3B the gap is 0.530 with an
interval of [0.462, 0.597]. The corresponding full-set trial-level gaps are
0.609 and 0.578. Both matched-subset intervals exclude zero, supporting a
class-change interpretation at similar measured edit magnitudes.

\subsection{Selective accuracy on the replacement}
Table~\ref{tab:follow} complements answer changes with class recognition.
Same-class replacement retains accuracies of 0.859, 0.841, and 0.795 for
Qwen2.5-Omni-7B, Qwen2.5-Omni-3B, and Fun-Audio-Chat-8B, respectively,
compared with 0.882, 0.853, and 0.847 on the originals. Same-class accuracy
remains within 5.2 percentage points of original accuracy. The matched rewrite
therefore retains recognition performance close to the original. On eligible
trials, soundscape-weighted selective accuracy for the inserted class is 0.819, 0.801, and
0.737. On these trials, the correct different-class answer names the inserted
class. This directional measure distinguishes recognition of the replacement
from answer movement alone, as local class changes often lead models to identify
the newly inserted event.

\section{Scope}
\label{sec:limitations}

We synthesize soundscapes to provide exact event boundaries, same-class
donors, and a background fixed across conditions. This enables matched local
replacement while limiting ecological realism. The evaluation covers 33 query
classes at one acoustic operating point. Selective accuracy is computed on
each model's eligible subset. Confidence intervals resample soundscapes. The 600 soundscapes use 458
background recordings, with up to eight soundscapes sharing one background. Extending the comparison to
natural recordings requires verified event intervals and suitable same-class
donors.

\section{Conclusion}
\label{sec:discussion}

Matched same-class rewriting provides a reference for interpreting answer
changes after a local sound-event replacement. With the edited span and
procedure held fixed, changing the queried class produces 49.7 to 60.9
percentage points more answer changes than preserving it across three models
from two architecture families. The gap is positive for all 33 query classes,
while off-target and sham edits elicit much smaller responses. On eligible
trials, soundscape-weighted selective accuracy of 73.7 to 81.9\% shows that
revised answers often identify the inserted class. The matched contrast and
directional check together provide a controlled way to test whether model
answers track the acoustic identity of a queried event.

\section*{Acknowledgments}
This work received no external funding. The authors have no competing interests
to disclose.

\section*{Compliance with Ethical Standards}
This study recruits no human participants and collects no personal data. The informal listening checks were performed by the authors themselves on synthesized audio. Sound events come from FSD50K through the FSD-MIX-SED source release, and backgrounds come from TUT Urban Acoustic Scenes 2018. Each corpus is used under its own terms, and no redistribution right is claimed for either.

\bibliographystyle{IEEEbib}
\bibliography{reference-refs}
\end{document}